\documentclass[aps,prd,twocolumn,superscriptaddress,floatfix,nofootinbib]{revtex4}

\usepackage{color}
\usepackage[dvips]{graphicx}

\newcommand{\lsim}{\mathrel{\mathop{\kern 0pt \rlap
      {\raise.2ex\hbox{$<$}}}\lower.9ex\hbox{\kern-.190em $ \sim$}}}

\newcommand{\gsim}{\mathrel{\mathop{\kern 0pt
      \rlap{\raise.2ex\hbox{$>$}}}\lower.9ex\hbox{\kern-.190em $\sim$}}}

\newcommand{\beq}{\begin{equation}}
\newcommand{\eeq}{\end{equation}}

\newcommand{\be}{\begin{equation}}
\newcommand{\ee}{\end{equation}}
\newcommand{\beqarr}{\begin{eqnarray}}
\newcommand{\eeqarr}{\end{eqnarray}}

\newcommand{\widesigmav}{\widetilde{\langle \sigma_{\rm ann} v \rangle}}
\newcommand{\sigmavint}{\langle \sigma_{\rm ann} v \rangle_{\rm int}}

\begin{document}

\title{Phenomenology of light neutralinos in view of recent results \\ at the CERN Large
Hadron Collider}

\thanks{Preprint number: DFTT 34/2011}


%
\author{A. Bottino}
\affiliation{Dipartimento di Fisica Teorica, Universit\`a di Torino \\
Istituto Nazionale di Fisica Nucleare, Sezione di Torino \\
via P. Giuria 1, I--10125 Torino, Italy}
\author{N. Fornengo}
\affiliation{Dipartimento di Fisica Teorica, Universit\`a di Torino \\
Istituto Nazionale di Fisica Nucleare, Sezione di Torino \\
via P. Giuria 1, I--10125 Torino, Italy}
\author{S. Scopel}
\affiliation{Department of Physics, Sogang University\\
Seoul, Korea, 121-742}

\date{\today}

\begin{abstract}

We review the status of the phenomenology of light neutralinos in an
effective Minimal Supersymmetric extension of the Standard Model
(MSSM) at the electroweak scale, in light of new results obtained at
the CERN Large Hadron Collider. First we consider the impact of the
new data obtained by the CMS Collaboration on the search for the Higgs
boson decay into a tau pair, and by the CMS and LHCb Collaborations on
the branching ratio for the decay $B_s \rightarrow {\mu}^{+} +
{\mu}^{-}$.  Then we examine the possible implications of the excess
of events found by the ATLAS and CMS Collaborations in a search for a
SM--like Higgs boson around a mass of 126 GeV, with a most likely mass
region (95\% CL) restricted to 115.5--131 GeV (global statistical
significance about 2.3 $\sigma$).  From the first set of data we
update the lower bound of the neutralino mass to be about 18 GeV. From
the second set of measurements we derive that the excess around
$m^{SM}_H$ = 126 GeV, which however needs a confirmation by further
runs at the LHC, would imply a neutralino in the mass range 18 GeV
$\lsim m_{\chi} \lsim$ 38 GeV, with neutralino--nucleon elastic cross
sections fitting well the results of the dark matter direct search
experiments DAMA/LIBRA and CRESST.

\end{abstract}

\pacs{95.35.+d,11.30.Pb,12.60.Jv,95.30.Cq}

\maketitle

%%%%%%%%%%%%%%%%%%%%%%%%%%%%%%%%%%%%%%%%%%%%%%%%%%%%%%%%%%%%%%%%%%%%%%%%%%%%%
\section{Introduction}
\label{sec:intro}

The phenomenology of light neutralinos has been thoroughly discussed
in Refs. \cite{lowneu,interpreting,discussing} within
an effective Minimal Supersymmetric extension of the
Standard Model (MSSM) at the electroweak (EW) scale, where the
usual hypothesis of gaugino--mass universality at the scale of Grand Unification (GUT)
of the SUGRA models is removed (this model containing neutralinos of mass
$m_{\chi} \lsim$ 50 GeV was dubbed Light Neutralino Model (LNM) \cite{discussing}); this
denomination will also be maintained here).

In Refs. \cite{lowneu,interpreting,discussing} it was shown that, in case of 
R--parity
conservation, a light neutralino within the LNM, when it happens to be the Lightest
Supersymmetric Particle (LSP), constitutes an extremely interesting
candidate for the dark matter in the Universe, with direct
detection rates accessible to experiments of present generation.
More specifically, the following results were obtained:
a) a lower bound on $m_{\chi}$ was derived from the  cosmological upper limit
on the cold dark matter density; b) it was shown that the population of light
neutralinos fits quite well the DAMA/LIBRA annual modulation results 
\cite{dama2004,dama2010} over a wide range
of $m_{\chi}$; c) this same population can explain also results of other direct searches
for dark matter (DM) particles which show positive results (CoGeNT \cite{cogent},
CRESST \cite{cresst}) or possible hints (two--event CDMS \cite{cdms1})
in some restricted intervals of $m_{\chi}$ \cite{noicdms,discussing,observ}. 

It is obvious that the features of the light neutralino population,
and its relevant properties (a-c), drastically depend on the
intervening constraints which follow from new experimental results. Of
particular impact over the details of the phenomenological aspects of
the LNM are the new data obtained at the CERN Large Hadron Collider
(LHC) which, in force of its spectacular performance, is providing a
profusion of new information.  In this respect the most relevant
results of LHC concern: i) the lower bounds on the squark and gluino
masses, ii) the correlated bounds on $\tan \beta$ (ratio of the two
Higgs v.e.v.'s) and $m_A$ (mass of the CP--odd neutral Higgs boson)
derived from the searches for neutral Higgs bosons into a tau--lepton
pair, iii) a new strict upper bound on the branching ratio for the
decay $B_s \rightarrow {\mu}^{+} + {\mu}^{-}$, iv) the indication of a
possible signal (at a statistical significance of 2.3 $\sigma$) for a
SM--like Higgs boson with a mass of about 126 GeV \cite{atlas,cms}.

The impact of item (i) on the LNM was already considered in
Ref. \cite{impact}.  In the present paper we derive the consequences
that the new bounds from searches for neutral Higgs bosons into a
tau--lepton pair and from $BR(B_s \rightarrow {\mu}^{+} + {\mu}^{-})$
(item (ii) and (iii) above) have on the phenomenology of the light
neutralinos and discuss the implications that a Higgs boson at about
126 GeV (item iv) could have, in case this preliminary experimental
indication is confirmed in next LHC runs.

\section{Features of the Light Neutralino Model}
\label{sec:features}

The LNM is an effective MSSM scheme at the electroweak scale, with the
following independent parameters: $M_1$, $M_2$, $M_3$, $\mu$,
$\tan\beta$, $m_A$, $m_{\tilde q_{12}}$, $m_{\tilde t}$, $m_{\tilde
  l_{12,L}}$, $m_{\tilde l_{12,R}}$, $m_{\tilde {\tau}_L}$, $m_{\tilde
  {\tau}_R}$ and $A$. We stress that the parameters are defined at the
EW scale.  Notations are as follows: $M_1$, $M_2$ and $M_3$ are the
U(1), SU(2) and SU(3) gaugino masses (these parameters are taken here
to be positive), $\mu$ is the Higgs mixing mass parameter, $\tan\beta$
the ratio of the two Higgs v.e.v.'s, $m_A$ the mass of the CP--odd
neutral Higgs boson, $m_{\tilde q_{12}}$ is a squark soft--mass common
to the squarks of the first two families, $m_{\tilde t}$ is the squark
soft--mass for the third family, $m_{\tilde l_{12,L}}$ and $m_{\tilde
  l_{12,R}}$ are the slepton soft--mass common to the L,R components
of the sleptons of the first two families, $m_{\tilde {\tau}_L}$ and
$m_{\tilde {\tau}_R}$ are the slepton soft--mass of the L,R components
of the slepton of the third family, $A$ is a common dimensionless
trilinear parameter for the third family, $A_{\tilde b} = A_{\tilde t}
\equiv A m_{\tilde t}$ and $A_{\tilde \tau} \equiv A (m_{\tilde
  {\tau}_L}+m_{\tilde {\tau}_R})/2$ (the trilinear parameters for the
other families being set equal to zero).  In our model, no gaugino
mass unification at a Grand Unified scale is assumed, and therefore
$M_1$ can be sizeably lighter than $M_2$.  Notice that the present
version of the LNM represents an extension of the model discussed in
our previous papers \cite{lowneu,interpreting,discussing}, where a
common squark and the slepton soft mass was employed for the 3
families.

The linear superposition of bino $\tilde B$, wino $\tilde W^{(3)}$
and of the two Higgsino states $\tilde H_1^{\circ}$, $\tilde
H_2^{\circ}$ which defines the neutralino state of lowest mass $m_{\chi}$ is written here as:
\begin{equation}
\chi \equiv a_1 \tilde B + a_2 \tilde W^{(3)} +
a_3 \tilde H_1^{\circ} + a_4  \tilde H_2^{\circ}.
\label{neutralino}
\end{equation}

\subsection{The cosmological bound}
\label{sec:cosmological}

Since no
gaugino--mass unification at a GUT scale is assumed in our LNM
(at variance with one of the major assumptions in mSUGRA), in this model
the neutralino mass is not bounded by the lower limit
$m_{\chi} \gsim$ 50 GeV that is commonly derived in mSUGRA schemes
from the LEP lower bound on the chargino mass (of about 100 GeV).
However, in the case of R--parity conservation the neutralino, when occurs to be the LSP, has a lower limit on its mass $m_{\chi}$
which can be derived from the cosmological upper bound on the
cold dark matter (CDM) relic abundance $\Omega_{CDM} h^2$.
Actually, by employing this procedure, in Ref. \cite{lowneu} a value
of 6--7 GeV for the lower limit of $m_{\chi}$ was obtained, and this value was subsequently updated to
the value of about 8 GeV  in Refs. \cite{discussing,observ} as derived
from the experimental data available at that time.
Now, with the advent of fresh data from LHC, the lower bound on
$m_{\chi}$ has to be redetermined; this will be done in 
Sect. \ref{sec:generic}).

To set the general framework, let us recall that
the neutralino relic abundance is given by:
\begin{equation}
\Omega_{\chi} h^2 = \frac{x_f}{{g_{\star}(x_f)}^{1/2}} \frac{9.9 \cdot
10^{-28} \; {\rm cm}^3 {\rm s}^{-1}}{\widesigmav},
\label{omega}
\end{equation}

\noindent
where $\widesigmav \equiv x_f \sigmavint$, $\sigmavint$ being the integral from the present temperature up to
the freeze--out temperature $T_f$ of the thermally averaged product of
the annihilation cross--section times the relative velocity of a pair
of neutralinos, $x_f$ is defined as $x_f \equiv m_{\chi}/T_f$ and
${g_{\star}(x_f)}$ denotes the relativistic degrees of freedom of the
thermodynamic
bath at $x_f$.  For $\widesigmav$
we will use the standard expansion in S and P waves:
$\widesigmav \simeq \tilde{a} + \tilde{b}/(2 x_f) $.
Notice that in the LNM no coannihilation effects are present in the
calculation of the relic abundance, due to the large mass splitting
between the mass of the neutralino ($m_{\chi}<50$ GeV) and those of
sfermions and charginos.

The annihilation processes which contribute to $\widesigmav$ at the
lowest order are: i) exchange of a Higgs boson in the s--channel, ii)
exchange of a sfermion in the t--channel, iii) exchange of the
Z--boson in the s--channel. In the physical region which we are going
to investigate, which entails light values for the masses of
supersymmetric Higgs bosons $m_h, m_A, m_H$ (for the lighter CP--even
$h$, the CP--odd $A$ and the heavier CP--even $H$, respectively) and a
light mass for the stau $\tilde{\tau}$, the contribution of the
Z--exchange is largely subdominant compared to the first two which can
be of the same order, with a dominance of the $A$--exchange
contribution for $m_{\chi} \lsim$ 28 GeV, and a possible dominance of
the $\tilde{\tau}$--exchange afterward (see numerical results in
Fig. \ref{fig3}).

In our numerical evaluations all relevant contributions to the pair
annihilation cross--section of light neutralinos are included.
However, an approximate expression for $\Omega_{\chi} h^2$, valid for
very light neutralinos proves very useful to obtain an analytic
formula for the lower bound for the neutralino mass.  Indeed, for
$m_{\chi} \lsim$ 28 GeV when $\widesigmav$ is dominated by the
$A$--exchange, $\Omega_{\chi} h^2$ may be written as \cite{lowneu}:

\begin{widetext}
\begin{eqnarray}
\Omega_{\chi} h^2 &\simeq& \frac{4.8 \cdot 10^{-6}}{\rm GeV^{2}} \frac{x_f}{{g_{\star}(x_f)}^{1/2}}
\frac{1}{ a_1^2 a_3^2 \tan^2\beta }
m_A^4
\frac{[1-(2m_{\chi})^2/m_A^2]^2}{m_{\chi}^2~[1-m_b^2/m_\chi^2]^{1/2}}
\frac{1}{(1 + \epsilon_b)^2},
\label{eq:omega0}
\end{eqnarray}
\end{widetext}
where $\epsilon_b$ is a quantity which enters in the relationship between the b--quark
running mass and the corresponding Yukawa coupling (see Ref. \cite{higgs} and references quoted therein).
For neutralino masses in the range $m_{\chi}$ = (10--20) GeV, ${g_{\star}(x_f)}^{1/2} \simeq$ 2.5.
In deriving this expression, one has taken into account that here
the following hierarchy holds for the coefficients $a_i$ of $\chi$ \cite{discussing}:
\begin{equation}
|a_1| > |a_3| \gg |a_2|, |a_4|,
\label{hierarchy1}
\end{equation}
\noindent
whenever $\mu/m_{\chi} \gsim$ a few. In  Ref. \cite{discussing} it is also shown that in this regime:
\begin{equation}
a_1^2 a_3^2 \simeq \frac{\sin^2 \theta_W \; m^2_Z \; \mu^2}{(\mu^2 + \sin^2 \theta_W \; m^2_Z)^2} \simeq  \frac{0.19 \; \mu^2_{100}}{(\mu^2_{100} + 0.19)^2},
\label{a1a3}
\end{equation}
\noindent where $\mu_{100}$ is $\mu$ in units of 100 GeV.  From this
formula and the LEP lower bound $|\mu| \gsim$ 100 GeV, we obtain
$(a_1^2 a_3^2)_{\rm max} \lsim 0.13$. This upper bound is essentially
equivalent to one which can be derived from the upper bound on the
width for the $Z$--boson decay into a light neutralino pair: $(a_1^2
a_3^2)_{\rm max} \lsim 0.12$ \cite{discussing}.

By imposing that the neutralino relic abundance does not exceed
the observed upper bound for cold dark matter (CDM), {\it i.e.}
$\Omega_{\chi} h^2 \leq (\Omega_{CDM} h^2)_{\rm max}$, we obtain
the following lower bound on the  neutralino mass:
\begin{widetext}
\begin{equation}
m_{\chi}~ \frac{[1-m_b^2/m_\chi^2]^{1/4}}{[1-(2m_{\chi})^2/m_A^2]}
\gsim 17 ~ {\rm GeV} \left(\frac{m_A}{90 \; {\rm GeV}} \right)^2 \left(\frac{15}{\tan \beta}\right)
\left(\frac{0.12}{a_1^2 a_3^2}\right)^{\frac{1}{2}}
\left(\frac{0.12}{(\Omega_{CDM} h^2)_{\rm max}}\right)^{\frac{1}{2}}.
\label{ma}
\end{equation}
\end{widetext}
Here we have taken as default value for
$(\Omega_{CDM} h^2)_{\rm max}$ the
numerical value which represents the 2$\sigma$ upper bound to $(\Omega_{CDM}
h^2)_{\rm max}$ derived from the results of Ref. \cite{wmap}. For
$\epsilon_b$ we have used a value which is representative of the typical range
obtained numerically in our model:
$\epsilon_b = -0.08$.

\subsection{Neutralino--nucleon elastic cross section}
\label{sec:elastic}

%%%---%%%
We turn now to the evaluation of the neutralino-nucleon elastic cross section
$\sigma_{\rm scalar}^{(\rm nucleon)}$, since we are interest here
in the comparison of our theoretical evaluations with the most recent data from
experiments of direct searches for DM particles.

Notice that we consider here only coherent neutralino--nucleus cross section, thus spin-dependent
couplings are disregarded, and the neutralino--nucleon cross section are derived from the coherent
neutralino--nucleus cross section in the standard way.

The neutralino--nucleon scattering then takes contributions from ($h, A, H$) Higgs boson exchange in the t--channel and from the squark exchange in the s-channel; the $A$--exchange contribution is suppressed by kinematic effects. In the supersymmetric parameter region considered in the present paper the contributions from the $h$ and $H$ exchanges are largely dominant over the squark exchange, with a sizable dominance of the
$h$ exchange over the $H$ one (a quantitative analysis of this point will be given in Sect. \ref{sec:generic} in connection with Fig.~\ref{fig4}). An approximate expression for $\sigma_{\rm scalar}^{(\rm nucleon)}$, valid at small
values of $m_{\chi}$, is obtained by including only the dominant contribution of the $h$ boson exchange \cite{discussing}:

\begin{widetext}
\begin{equation}
\sigma_{\rm scalar}^{(\rm nucleon)} \simeq 9.7  \times 10^{-42} \; {\rm cm^2} \;
\left(\frac{a_1^2 a_3^2}{0.13} \right)
\left(\frac{\tan \beta}{15} \right)^2
\left(\frac{90 \; {\rm GeV}}{m_h} \right)^4
\left(\frac{g_d}{290 ~ {\rm MeV}} \right)^2.
\label{sel2}
\end{equation}
\end{widetext}

\noindent
where

\begin{equation}
g_d \equiv [m_d \langle N|\bar{d} d |N\rangle + m_s \langle N|\bar{s} s |N\rangle +
m_b \langle N|\bar{b} b |N\rangle].
\label{eq:gd}
\end{equation}
\noindent and the matrix elements $\langle N|\bar{q}q|N \rangle$
denote the scalar quark densities of the $d,s,b$ quarks inside the
nucleon.

In Eq. (\ref{sel2}) we have used as {\it reference} value for $g_d$
the value $g_{d,\rm ref} = 290$ MeV employed in our previous papers
\cite{interpreting,discussing}. We recall that this quantity is affected by large
uncertainties \cite{uncert2} with $\left({g_{d,\rm max}}/{g_{d,\rm
    ref}}\right)^2 = 3.0$ and $\left({g_{d,\rm min}}/{g_{d,\rm
    ref}}\right)^2 = 0.12$ \cite{interpreting,discussing}.  Notice that these
uncertainties still persist \cite{dinter,martin}.  Our reference value
$g_{d,\rm ref} = 290$ MeV is larger by a factor 1.5 than the central
value of Ref. \cite{efo}, frequently used in the literature.

By employing Eq. (\ref{eq:omega0}) and Eq. (\ref{sel2}) we find that
any neutralino configuration, {\it whose relic abundance stays in the
  cosmological range for CDM} ({ \it i.e.} $(\Omega_{CDM} h^2)_{\rm
  min} \leq \Omega_{\chi} h^2 \leq (\Omega_{CDM} h^2)_{\rm max}$ with
$(\Omega_{CDM} h^2)_{\rm min} = 0.098$ and $(\Omega_{CDM} h^2)_{\rm
  max} = 0.12$) and passes all particle--physics constraints, has an
elastic neutralino--nucleon cross--section given approximately by
\cite{discussing}:

\begin{widetext}
\begin{equation}
\sigma_{\rm scalar}^{(\rm nucleon)} \simeq (2.7 - 3.4) \times 10^{-41} \; {\rm cm^2}  \;
\left(\frac{g_d}{290 ~ {\rm MeV}} \right)^2
\left(\frac{m_A}{m_h}\right)^4
\frac{[1-(2m_{\chi})^2/m_A^2]^2}{(m_{\chi}/(10 \; {\rm GeV})^2 \; [1- m_b^2/m_\chi^2]^{1/2}}.
\label{bound}
\end{equation}
\end{widetext}

\noindent
Notice that in the range 90 GeV $\leq m_A \leq$ 120 GeV the maximal
values of the ratio $m_A/m_h$ are of order one within a few percent
(see left panel of Fig. \ref{fig2}).

We recall that for neutralino configurations whose relic abundance stays below the
cosmological range for CDM, {\it i.e.} have $\Omega_{\chi} h^2 < (\Omega_{CDM} h^2)_{\rm min}$
one has to associate to $\sigma_{\rm scalar}^{(\rm nucleon)}$ a local density rescaled by a factor
$\xi = \rho_{\chi} /
\rho_0$,  as compared to the
total local DM density $\rho_{\chi}$; $\xi$ is conveniently taken as $\xi = {\rm min}\{1, \Omega_{\chi}
h^2/(\Omega_{CDM} h^2)_{\rm min}\}$ \cite{gaisser}.

Furthermore, we note that Eq. (\ref{bound}) is valid when the $A$
boson exchange is dominating in the neutralino pair annihilation
process (in the s--channel). As mentioned above, this occurs for
$m_{\chi} \lsim$ 28 GeV.  For higher neutralino masses the actual
values of $\sigma_{\rm scalar}^{(\rm nucleon)}$ are somewhat higher
than those provided by Eq. (\ref{bound}).

\subsection{Constraints}
\label{sec:constraints}

%%%%%%%%%%%%%%%%%%%%%%%%%%%%%%%%%%%%%%%%%%%%%%%%%
\begin{figure}[t]
\includegraphics[width=\columnwidth]{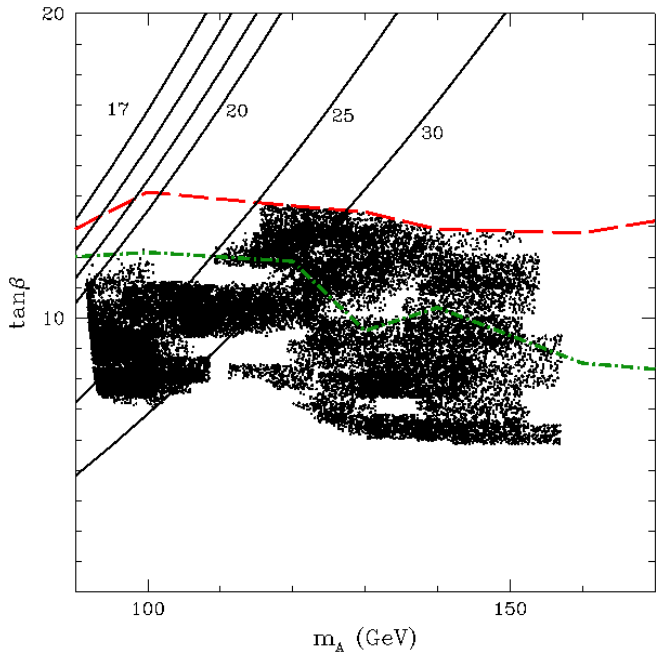}
  \caption{Upper bounds in the $m_A$ -- $\tan\beta$ plane, derived
    from searches of the neutral Higgs decays into a tau pair at the
    LHC.  The dot--dashed line denotes the 95\% CL upper bound
    reported in Ref. \cite{cms029}. The dashed line displays the
    expected upper bound (in case of no positive signal for an
    integrated luminosity of 3 fb$^{-1}$) as evaluated in
    Refs. \cite{baglio1,baglio2}. The scatter plot denotes
    configurations of the LNM. The solid lines (some of which labeled
    by numbers) denote the cosmological bound $\Omega_\chi h^2 \leq
    (\Omega_{CDM} h^2)_{\rm max}$ for a neutralino whose mass is given
    by the corresponding number (in units of GeV), as obtained by
    Eqs. (\ref{ma}), with $\epsilon_b = -0.08$ and $(\Omega_{CDM}
    h^2)_{\rm max}=0.12$. For any given neutralino mass, the allowed
    region is above the corresponding line.}
\label{fig1}
\end{figure}

%%%%%%%%%%%%%%%%%%%%%%%%%%%%%%%%%%%%%%%%%%%%%%%%%

%%%%%%%%%%%%%%%%%%%%%%%%%%%%%%%%%%%%%%%%%%%%%%%%%
\begin{figure*}[t]
\includegraphics[width=\columnwidth]{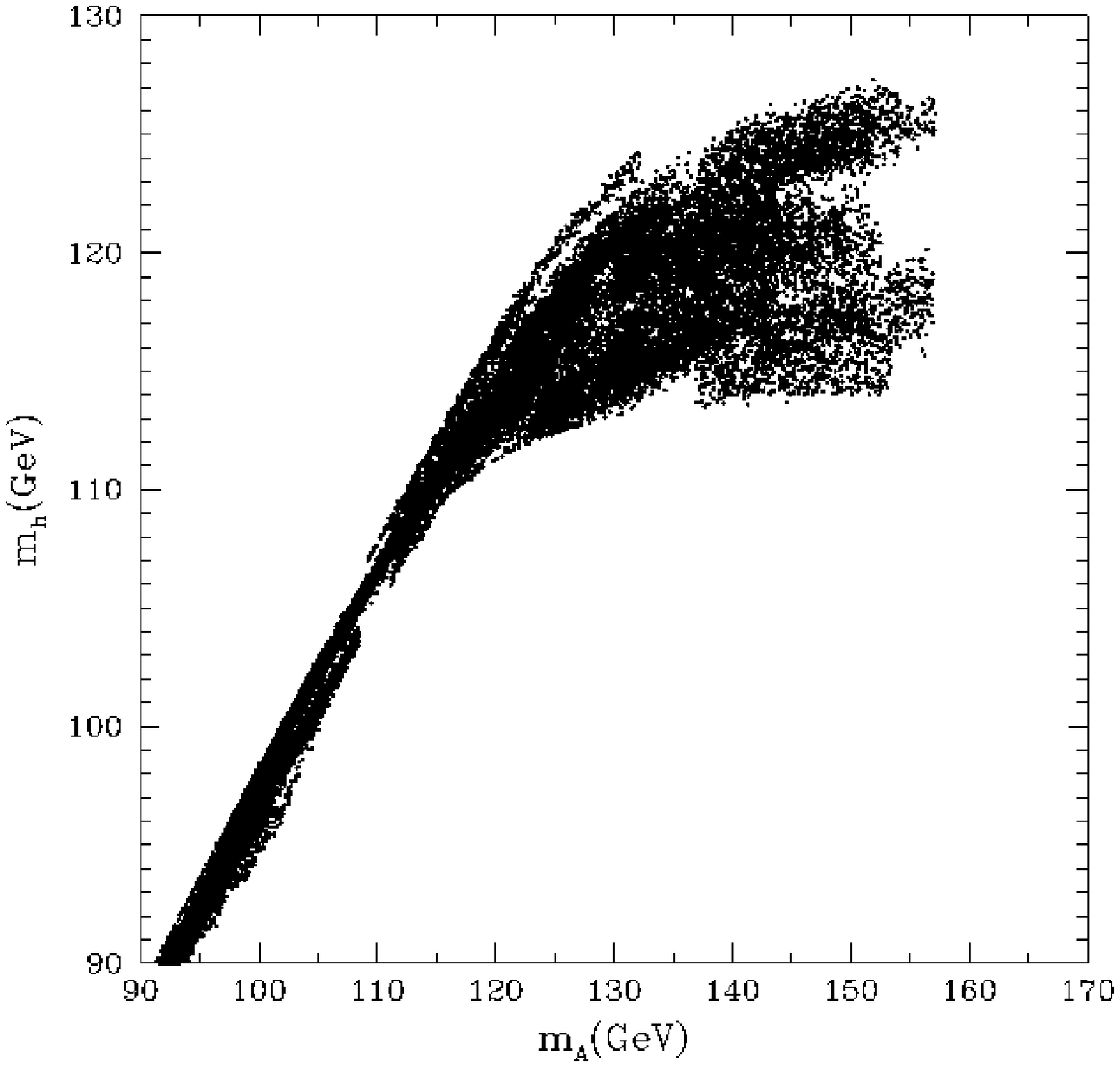}
\includegraphics[width=\columnwidth]{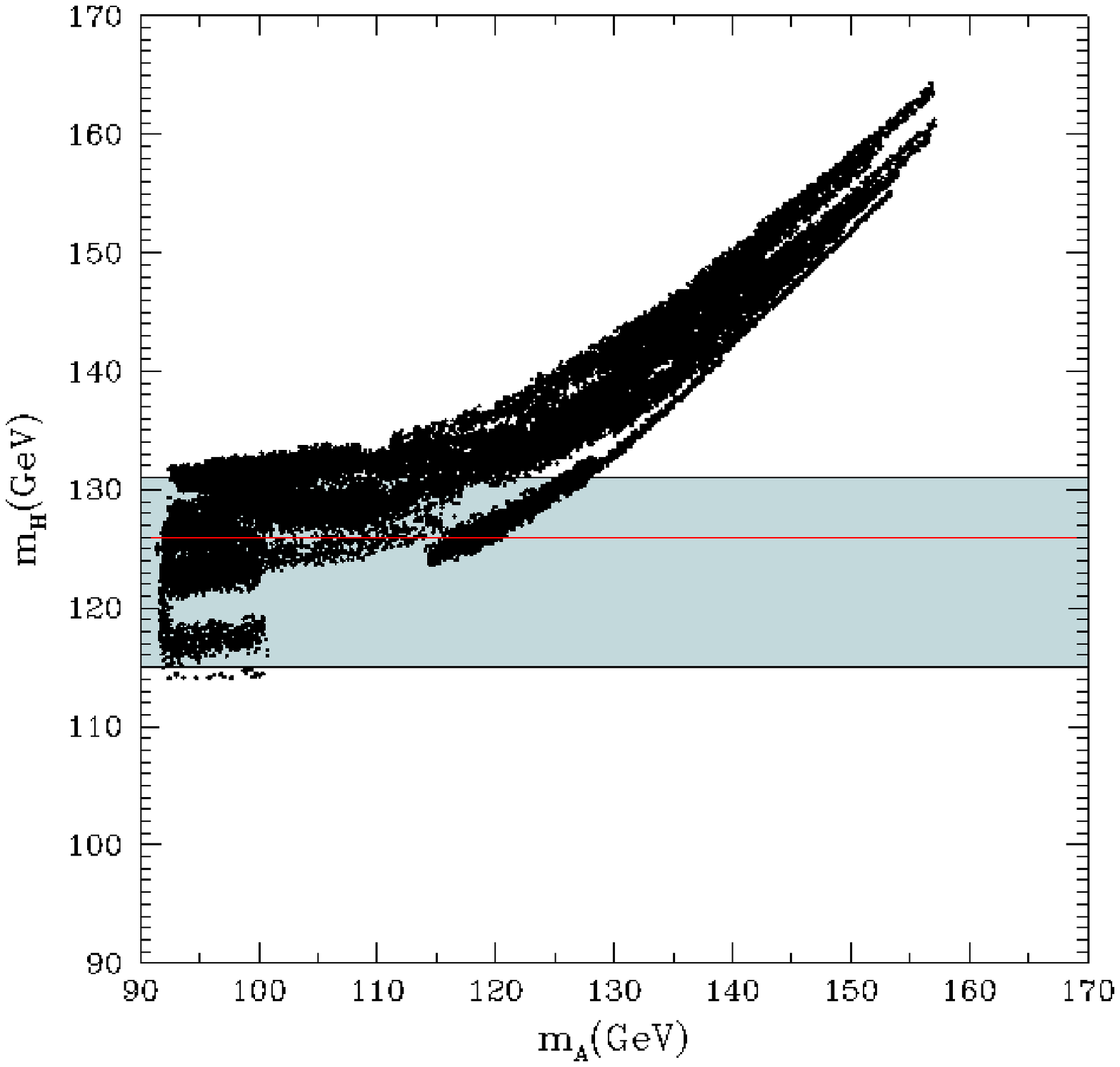}
  \caption{Relation among the Higgs masses in the LNM. In the left
    panel, the correlation between $m_h$ and $m_A$ is shown. In the
    right panel, the correlation between $m_H$ and $m_A$ is given. The
    horizontal (red) line and the shaded band around it denote the
    value of 126 GeV for the Higgs mass (and the 95 \% CL region
    between 115.5 GeV and 131 GeV) compatible with the excess of
    events observed by ATLAS \cite{atlas} and CMS \cite{cms}.}
\label{fig2}
\end{figure*}

%%%%%%%%%%%%%%%%%%%%%%%%%%%%%%%%%%%%%%%%%%%%%%%%%

To single out the physical supersymmetric configurations within our
LNM the following experimental constraints are imposed: accelerators
data on supersymmetric and Higgs boson searches at the CERN $e^+ e^-$
collider LEP2 \cite{LEPb}; the upper bound on the invisible width for
the decay of the $Z$--boson into non Standard Model particles:
$\Gamma(Z \rightarrow \chi \chi) <$ 3 MeV \cite{aleph05,pdg};
measurements of the $b \rightarrow s + \gamma$ decay process
\cite{bsgamma}: 2.89 $\leq BR(b \rightarrow s \gamma) \cdot 10^{4}
\leq$ 4.21 is employed here (this interval is larger by 25\% with
respect to the experimental determination \cite{bsgamma} in order to
take into account theoretical uncertainties in the supersymmetric
(SUSY) contributions \cite{bsgamma_theorySUSY} to the branching ratio
of the process (for the Standard Model calculation, we employ the NNLO
results from Ref. \cite{bsgamma_theorySM})); the measurements of the
muon anomalous magnetic moment $a_\mu \equiv (g_{\mu} - 2)/2$: for the
deviation, $\Delta a_{\mu} \equiv a_{\mu}^{\rm exp} - a_{\mu}^{\rm
  the}$, of the experimental world average from the theoretical
evaluation within the Standard Model we use here the (2 $\sigma$)
range $31 \leq \Delta a_{\mu} \cdot 10^{11} \leq 479 $, derived from
the latest experimental \cite{bennet} and theoretical \cite{davier}
data (the supersymmetric contributions to the muon anomalous magnetic
moment within the MSSM are evaluated here by using the formulae in
Ref. \cite{moroi}); the search for charged Higgs bosons in top quark
decay at the Tevatron \cite{abazov_top}; the recently improved upper
bound (at $95 \%$ C.L.) on the branching ratio for the decay $B_s
\rightarrow {\mu}^{+} + {\mu}^{-}$: $BR(B_s \rightarrow {\mu}^{+}
{\mu}^{-}) < 1.08 \times 10^{-8}$ \cite{cmslhcb} (see also
Refs. \cite{cms_bsmumu,lhcb_bsmumu}) and the constraints related to
$\Delta M_{B,s} \equiv M_{B_s} - M_{\bar{B}_s}$
\cite{buras_delta,isidori}.

A further bound, which plays a most relevant role in constraining the
supersymmetric parameter space, is represented by the results of
searches for Higgs decay into a tau pair.  Indeed colliders have a
good sensitivity to the search for decays ($\phi \rightarrow b
\bar{b}$ or $\phi \rightarrow \tau \bar{\tau}$) (where $\phi = h, A,
H$) in the regime of small $m_A$ and large $\tan \beta$, because in
this region of the supersymmetric parameters the couplings of one of the
neutral Higgs bosons to the down--fermions are enhanced
\cite{bernardi}. This experimental investigation was thoroughly
carried out at the Tevatron and is now underway at the LHC. No signal
for these decays has been found so far, thus successive measurements
have progressively disallowed substantial regions in the
supersymmetric parameters space at small $m_A$ and large $\tan \beta$.

 However, at present the actual forbidden region is not yet firmly
 established. The most stringent bounds provided in the $m_A$ -- $\tan
 \beta$ plane are reported by the CMS Collaboration in a preliminary
 form in Ref.~\cite{cms020} and Ref.~\cite{cms029}. The first report
 refers to a luminosity of 1.6 fb$^{-1}$, the second one to a
 luminosity of 4.6 fb$^{-1}$. It is worth noting that, in the range 90
 GeV $\leq m_{\chi} \leq$ 120 GeV, the bound on $\tan \beta$ given in
 the second report is less stringent than the limit given in the first
 one by a factor of $(20-40)\%$.  This circumstance suggests to take
 the present constraints with much caution.  A conservative attitude
 is also suggested by the considerations put forward in
 Refs. \cite{baglio1,baglio2} about the actual role of uncertainties
 in the derivation of the present bounds.

In Fig.~\ref{fig1}, which displays the plane $m_A$ -- $\tan \beta$, we
summarize the present situation as far as the constraints from the
collider searches for the neutral Higgs decays into a tau pair are
concerned. The dot--dashed line denotes the 95\% CL upper bound
reported in Ref. \cite{cms029}, accounting for a +1$\sigma$
theoretical uncertainty. The dashed line displays the expected upper
bound (in case of no positive signal for an integrated luminosity of 3
fb$^{-1}$) as evaluated in Refs. \cite{baglio1,baglio2}. We do not
mean to attribute to this expected bound the meaning of the most
realistic upper limit; we just take it as indicative of a conservative
estimate of the bound, and thus as a reasonable upper extreme of the
physical range to consider in our scan of the parameter space.

Notice that the regime of small $\tan \beta$ values is also compatible
with one of the physical regions selected by the branching ratio BR($B
\rightarrow \tau + \nu$) (see Fig.16 of Ref.~\cite{discussing}).

Also shown in Fig. \ref{fig1} are the curves which correspond to a
fixed value of $m_{\chi}$; these are calculated from Eq.~(\ref{ma}) by
replacing the inequality with an equality symbol and setting, for
definiteness, to 1 the two last factors of the
right--hand--side. Thus, for configurations with different values of
$(a_1^2 a_3^2)^{1/2}$, the $m_{\chi}$ value associated to each isomass
curve has to be scale up by the factor $(a_1^2 a_3^2/0.12)^{1/2}$.
The features of the scatter plot displayed in Fig. \ref{fig1} and its
implications will be discussed in the next Section.

We recall that also the cosmological constraint $\Omega_{\chi} h^2
\leq (\Omega_{CDM} h^2)_{\rm max}$, discussed in
Sect. \ref{sec:cosmological}, is implemented in our analysis.

The viability of very light neutralinos in terms of various  constraints
from collider data, precision observables and rare meson decays is also
considered in Ref. \cite{Dreiner:2009ic}. Perspectives for investigation of
these neutralinos at LHC  are analyzed in Ref. \cite{lhc1,lhc2} and
prospects for a very accurate mass measurement at ILC in Ref.
\cite{Conley:2010jk}.

\section{Results}
\label{sec:results}

According to the considerations developed up to now, it is clear that,
in order to examine the physical region relevant for light neutralinos
with a sizable elastic neutralino--nucleon cross section efficiently,
one has to set up a scan of the supersymmetric parameter space focused
on low values of $M_1$, restricted ranges of $m_A$ and $\mu$ close to
their minimal values as allowed by present experimental lower bounds,
and a range of $\tan \beta$ delimited from above by the bounds from
the neutral Higgs decays into a tau pair and from $BR(B_s \rightarrow
{\mu}^{+} + {\mu}^{-})$. The LEP limits on $\tan \beta$ and $m_A$ are
taken into account through the bounds derived from the
Higgs--strahlung of the Z--boson \cite{LEPb}.  The selection of the
parameters ranges has also to allow small values of the tau slepton,
to take care of the cosmological bound for neutralinos with $m_{\chi}
\gsim$ 28 GeV (see previous discussion in
Sect. \ref{sec:cosmological}).

For these reasons the scan of the parameter space adopted in the present paper is the following:
$1 \leq \tan \beta \leq 15$,
$100 \, {\rm GeV} \leq \mu \leq 200 \, {\rm GeV}$,
$10 \, {\rm GeV} \leq M_1 \leq 100 \, {\rm GeV}$,
$100 \, {\rm GeV} \leq M_2 \leq 2000 \, {\rm GeV}$,
$700 \, {\rm GeV} \leq m_{\tilde q_{12}} \leq 2000 \, {\rm GeV }$,
$100 \, {\rm GeV} \leq m_{\tilde t} \leq 1000 \, {\rm GeV }$,
$70 \, {\rm GeV} \leq  m_{\tilde l_{12,L}}, m_{\tilde l_{12,R}}, m_{\tilde {\tau}_L}, m_{\tilde {\tau}_R} \leq 150 \, {\rm GeV }$,
$90\, {\rm GeV }\leq m_A \leq 160 \, {\rm GeV }$,
$0.5 \leq A \leq 3$.

We turn now to the discussion of the physical results as obtained by
our numerical scans of the supersymmetric parameter space. First we
analyze the generic population of light neutralinos within the LNM
which takes into account all the constraints listed in the previous
Sect. \ref{sec:constraints}, then we will discuss the impact of the
excess seen by the ATLAS and CMS Collaborations at the LHC.

\subsection{The light neutralino population within the LNM}
\label{sec:generic}

A first result of our scans is shown in Fig. \ref{fig1}. From the
scatter plot displayed here one sees that the lower bound on the
neutralino mass turns out to be about 18 GeV. The depopulation in the
domain with $\tan \beta \gsim$ 12 and 90 GeV $\lsim m_A \lsim$ 100 GeV
with respect to our previous analyses \cite{discussing} is due to the
new bound $BR(B_s \rightarrow {\mu}^{+} {\mu}^{-}) < 1.08 \times
10^{-8}$ \cite{cmslhcb}.

In Fig. \ref{fig2} we display the correlation between $m_A$ and $m_h, m_H$, because this will be useful for the discussions to follow. From the
left panel of Fig. \ref{fig2} one can derive the values of the ratio $m_A/m_h$ which enters into the approximate estimate of the neutralino--nucleon elastic cross section due to the $h$--exchange contribution (see Eq. (\ref{bound}).

\begin{figure}[t]
\includegraphics[width=\columnwidth]{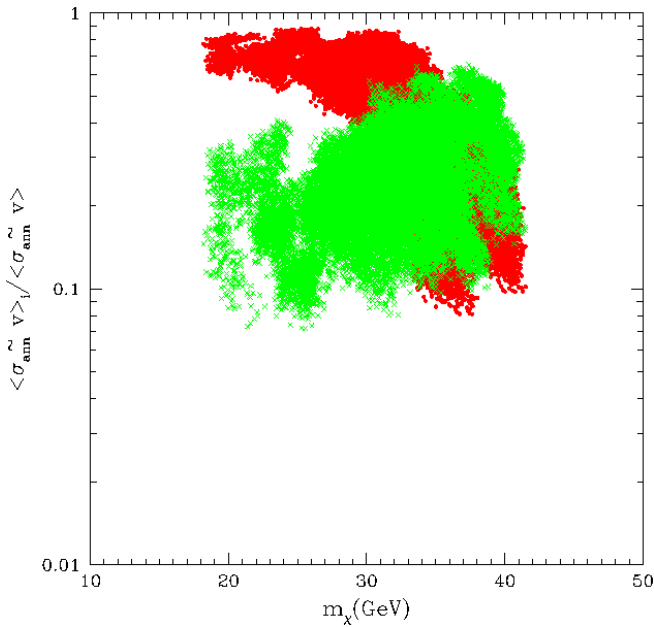}
  \caption{Fractional relevance of channels in the neutralino
    self--annihilation cross section $\widesigmav$ appearing in
    Eq. (\protect\ref{omega}) as a function of the neutralino mass in
    the LNM. The (red) points refer to annihilation through
    $A$--exchange; (green) crosses to annihilation through
    $\tilde\tau$--exchange.}
\label{fig3}
\end{figure}

\begin{figure}[t]
\includegraphics[width=\columnwidth]{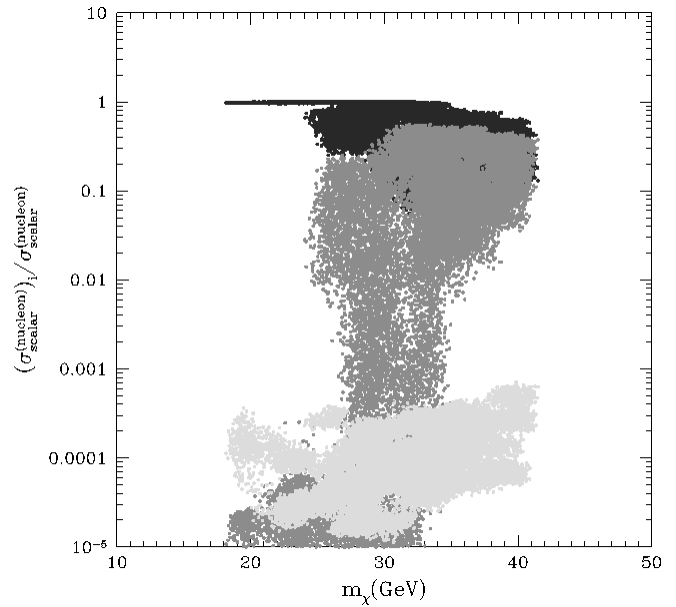}
  \caption{Fractional relevance of channels in the neutralino--nucleon elastic--scattering  
  cross section as a function of the neutralino mass in the LNM. From darker to lighter points: $h$--exchange, $H$--exchange, $\tilde q$--exchange.}
\label{fig4}
\end{figure}

Figs. \ref{fig3} and \ref{fig4} give the size of the various channels
contributing to the neutralino pair annihilation and to the
neutralino--nucleon elastic cross section, respectively. From
Fig. \ref{fig3} we observe in the neutralino pair annihilation cross
section a dominance of the $A$--exchange contribution for $m_{\chi}
\lsim$ 28 GeV, and a possible dominance of the
$\tilde{\tau}$--exchange for larger values of $m_{\chi}$, as
anticipated in Sect. \ref{sec:cosmological} (the contribution of the
$Z$--exchange is largely subdominant compared to the other two and is
not shown). Fig.~\ref{fig4} shows that in the direct detection cross
section the contributions from the $h$ and $H$ exchanges are largely
dominant over the squark exchange, with a sizable dominance of the $h$
exchange over the $H$ one.

%%%%%%%%%%%%%%%%%%%%%%%%%%%%%%%%%%%%%%

%%%%%%%%%%%%%%%%%%%%%%%%%%%%%%%%%%%%%%

%%%%%%%%%%%%%%%%%%%%%%%%%%%%%%%%%%%%%%
%%%%%%%%%%%%%%%%%%%%%%%%%%%%%%%%%%%%%%

 The scatter plot for the
quantity relevant for the comparison with the direct detection experimental results, $\xi \sigma_{\rm scalar}^{(\rm nucleon)}$,
is displayed in Fig.~\ref{fig5}. It is noticeable that our population of light neutralinos fits quite well a region
of compatibility of the DAMA/LIBRA data with the CRESST results in the
$m_{\chi}$--$\xi \sigma_{\rm scalar}^{(\rm nucleon)}$ plane.

Some comments are in order here:

a) The scatter plot shown in Fig.~\ref{fig5} is obtained with a
specific set of values for the hadronic quantities which establish the
coupling between the Higgs boson and the nucleon ({\it i.e.}  $g_{d} =
g_{d,\rm ref} = 290$ MeV). As mentioned in Sect.\ref{sec:elastic}, the
quantity $g_{d}$ suffers from large uncertainties \cite{efo}, so that
the scatter plot of Fig.~\ref{fig5} could actually move upwards by a
factor 3 or downward by a factor 0.12.

b) The experimental region of each individual experiment is sizeably
affected by uncertainties due to the estimate of the quenching
factor. In the case of the DAMA/LIBRA experiment the two regions are
illustrative (but not exhaustive) of the large effect introduced by
different evaluations of this factor \cite{observ}.

c) The position of the experimental regions $m_{\chi}$--$\xi
\sigma_{\rm scalar}^{(\rm nucleon)}$ strongly depends also on the DM
galactic distribution function (DF) employed in deriving these regions
from the experimental rates. Thus, their location relative to the
theoretical scatter plot changes depending on the galactic DM
properties \cite{Belli:1999nz}.  The domains shown in Fig.~\ref{fig5}
were obtained by using for the DF the standard isothermal sphere with
$\rho_0 = 0.30$ GeV cm$^{-3}$, $v_0 = 220$ km sec$^{-1}$, with $v_{\rm
  esc} = 650$ km sec$^{-1}$ for DAMA/LIBRA experiment and $v_{\rm esc}
= 544$ km sec$^{-1}$ for CRESST. The use of a DF with a larger
(smaller) value of $\rho_0$ would move downward (upward) the
experimental regions by a factor proportional to $\rho_0$. Increasing
(decreasing) the speeds generically produce a displacement towards
lower (higher) masses \cite{Belli:1999nz}.

In conclusion, by taking into account various sources of
uncertainties, mainly the ones mentioned in the two last items, the
experimental regions shown in Fig.~\ref{fig5} may change sizably. In
the case of the DAMA/LIBRA experiment the regions which encompass the
effects of various uncertainties are plotted in Figs. 1--3 and 7 of
Ref. \cite{observ}.

Negative results reported by other experiments of DM direct detection
\cite{xenon,cdms,kims} are in tension with the signals measured by
DAMA/LIBRA and CRESST. It should however be noted that a number of
questions about various physical and technical features of the
specific detectors or of the relevant analyses have been raised
\cite{dama1,collar1,collar2}. One further experiment, CoGeNT
\cite{cogent}, reports the measurement of an yearly--modulated
signal. If interpreted in terms of a coherently interacting dark
matter particle, this signal gives a region in the $m_{\chi}$--$\xi
\sigma_{\rm scalar}^{(\rm nucleon)}$ plane, which is approximately
located around $m_{\chi} \sim$ 10 GeV and $\xi \sigma_{\rm
  scalar}^{(\rm nucleon)} \sim (3-10) \times 10^{-41}$ cm$^2$, thus
somewhat displaced from the region singled out by the scatter plot of
Fig.6. However, a redetermination of the region towards smaller
$\sigma_{\rm scalar}^{(\rm nucleon)}$ and larger $m_{\chi}$ is being
undertaken by the CoGeNT Collaboration \cite{cogenttaup}.

\subsection{The neutralino subpopulation singled out by a Higgs at 126 GeV}
\label{sec:subpop}

%%%%%%%%%%%%%%%%%%%%%%%%%%%%%%%%%%%%%%
\begin{figure}[t]
\includegraphics[width=\columnwidth]{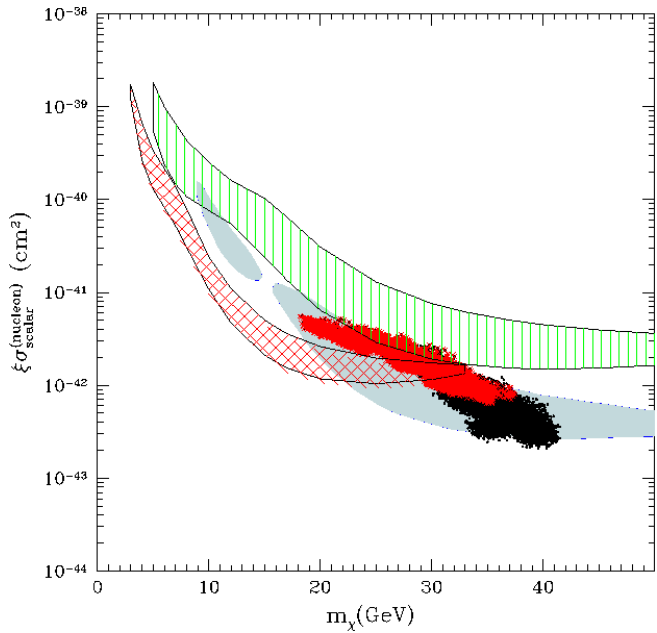}
  \caption{Neutralino--nucleon cross section $\xi \sigma_{\rm
      scalar}^{(\rm nucleon)}$ as a function of the neutralino mass
    for the LNM scan and for $g_{d,\rm ref}$ = 290 MeV.  The (red)
    crosses denote configurations with a heavy Higgs mass in the range
    compatible with the ATLAS \cite{atlas} and CMS \cite{cms} excess
    at the LHC.  The shaded areas denote the DAMA/LIBRA annual
    modulation regions: the upper area (vertical shade; green) refers
    to the case where constant values of 0.3 and 0.09 are taken for
    the quenching factors of Na and I, respectively\cite{observ}; the
    lower area (cross hatched; red) is obtained by using the
    energy--dependent Na and I quenching factors as established by the
    procedure given in Ref. \cite{tretyak}. The gray regions are those
    compatible with the CRESST excess \cite{cresst}. In all cases a
    possible channeling effect is not included.The halo distribution
    functions used to extract the experimental regions are given in
    the text.}
\label{fig5}
\end{figure}
%%%%%%%%%%%%%%%%%%%%%%%%%%%%%%%%%%%%%%

We turn now to the analysis of a subset of the neutralino population
considered in the previous section which would be selected by an
indication of a possible Higgs signal at the LHC. Actually, the ATLAS
Collaboration, in a search for a SM Higgs boson, measures an excess of
events around a mass of 126 GeV, and restricts the most likely mass
region (95 \% CL) to 115.5--131 GeV (global statistical significance
about 2.3 $\sigma$) \cite{atlas}. Similar results (with a lower
statistical significance) are presented by CMS \cite{cms}.  We address
the question of what might be the implications of these measurements
(in case the effect is confirmed in next runs at the LHC) under the
hypothesis that this possible signal is attributed to the production
of the heavier neutral CP--even Higgs boson $H$ of the MSSM
\cite{hein}.

Within our light neutralino population we select the subset of
configurations with 115 GeV $\leq M_H \leq$ 131 GeV. These are
contained in the band shown in the right panel of Fig.~\ref{fig2},
with values of the $m_A$ parameter in the range 90 GeV $\leq M_A \leq$
129 GeV. This subpopulation of light neutralinos would have a
neutralino--nucleon elastic cross section in the domain depicted in
Fig.~\ref{fig5} by (red) crosses, and would then be in amazing
agreement with the results of DM direct detection.

\begin{figure}[t]
\includegraphics[width=\columnwidth]{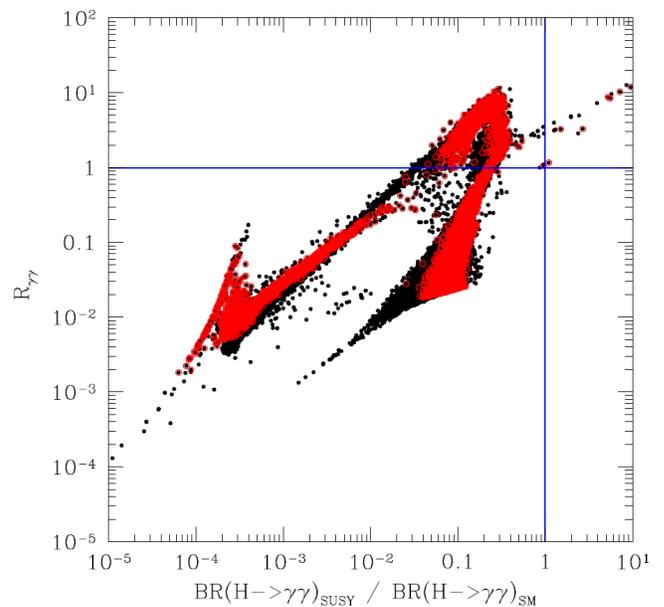}
  \caption{Production cross section ratio $R_{\gamma\gamma}\equiv
    [\sigma(gg\rightarrow H) \times BR(H\rightarrow
    \gamma\gamma)]_{MSSM}/[\sigma(gg\rightarrow h) BR(h\rightarrow
    \gamma\gamma)]_{SM}$ as a function of $BR(H\rightarrow
    \gamma\gamma)_{MSSM}/BR(h \rightarrow \gamma\gamma)_{SM}$ for the
    configurations discussed in Section \protect\ref{sec:generic}. Black points refer to $H$ masses
    in the range 115 GeV $\leq m_H \leq$ 131 GeV, while (red) circles refer to a $H$ mass interval
    more focussed around 126 GeV (specifically: 125 GeV $\leq m_H \leq$ 127 GeV).
}
\label{fig6}
\end{figure}

  The identification of a putative Higgs boson with the $H$ boson
  appears to be compatible in terms of production cross section and
  branching ratios. This is shown in Fig. \ref{fig6}, where the
  exclusive production cross section ratio $R_{\gamma\gamma}\equiv
  [\sigma(gg\rightarrow H) \times BR(H\rightarrow
  \gamma\gamma)]_{MSSM}/[\sigma(gg\rightarrow H) BR(H\rightarrow
  \gamma\gamma)]_{SM}$ is plotted as a function of $BR(H\rightarrow
  \gamma\gamma)_{MSSM}/BR(H \rightarrow \gamma\gamma)_{SM}$ for our
  configurations. Here $\sigma(gg\rightarrow H)$ is the Higgs
  production cross section through the gluon fusion process. We have
  calculated both quantities using FeynHiggs 2.8.6
  \cite{feynhiggs}. Indeed our population of light neutralinos
  contains many configurations which are in agreement with the
  putative Higgs signal. This is a property arising spontaneusly in
  our scenario. Notice that although the BR of Higgs decay into 2
  photons is typically smaller that the corresponding SM branchig
  ratio, $R_{\gamma\gamma}$ can be SM--like, due to enhanced
  production cross sections.

       Though imposing the above requirement would imply some further
       selection within the neutralino population previously discussed,
       we do not find in our scan any significant correlation between
       $R_{\gamma\gamma}$ and the properties of relic neutralinos, such
       as the neutralino relic abundance $\Omega_{\chi} h^2 $ or the
       neutralino--nucleon cross section $\xi \sigma_{\rm scalar}^{(\rm
       nucleon)}$. In fact $R_{\gamma\gamma}$ is mainly affected by the
       production cross section $\sigma(gg\rightarrow H)$, which depends
       on SUSY--QCD parameters that do not enter directly into the
       calculation of relic neutralino observables. Although a thorough
       analysis of these aspects is beyond the scope of the present
       paper, the previous considerations are sufficient to conclude
       that our scenario can be compatible with the possible Higgs
       signal at the LHC in a natural way.

\section{Conclusions}
\label{sec:conclusions}

We have reviewed the status of the phenomenology of light neutralinos
in an effective Minimal Supersymmetric extension of the Standard Model
(MSSM) at the electroweak scale, in light of new results obtained at
the CERN Large Hadron Collider. First we considered the impact of the
new data obtained by the CMS Collaboration on the search for the Higgs
boson decay into a tau pair, and by the CMS and LHCb Collaborations on
the branching ratio for the decay $B_s \rightarrow {\mu}^{+} +
{\mu}^{-}$, and we established that, on the basis of these data, the
new value for the lower bound of the neutralino mass is $m_{\chi}
\simeq$ 18 GeV.

Then we have examined the possible implications
of the excess of events found by the ATLAS and CMS Collaborations in a search for a SM--like
Higgs boson around a mass of 126 GeV,
with a most likely mass region (95 \% CL) restricted to 115.5--131 GeV (global statistical significance
about 2.3 $\sigma$).
 We have derived that
the excess around $m^{SM}_H$ = 126 GeV, which nevertheless needs a confirmation by further runs at the LHC,
 would imply
a neutralino in the mass range  18 GeV $\lsim m_{\chi} \lsim$ 38 GeV, with neutralino--nucleon elastic
cross sections
fitting well the results of the dark matter direct search experiments DAMA/LIBRA and CRESST.

It is worth stressing that light neutralinos in the mass range
considered here do not appear to be constrained by DM indirect
searches (such as astrophysical gamma fluxes of diffuse extragalactic
origin or from dwarf galaxies, and the low--energy cosmic antiproton
flux).  A detailed investigation of these aspects would however
deserve a dedicated analysis.

\medskip
\section*{Note Added}
After the submission of the present paper, a new upper bound (at $95 \%$
C.L.) on the branching ratio for the decay $B_s \rightarrow {\mu}^{+} +
{\mu}^{-}$ has been presented by the LHCb Collaboration: $BR(B_s
\rightarrow {\mu}^{+} {\mu}^{-}) < 4.5 \times 10^{-9}$ \cite{LHCb_2012}.
If the previous upper bound $BR(B_s \rightarrow {\mu}^{+} {\mu}^{-}) <
1.08 \times 10^{-8}$, employed in our analysis, is
replaced by the new LHCb upper limit, the lower bound on the neutralino
mass rises from the value of about 18 GeV, presented above, to
about 20 GeV.

\acknowledgments

A.B. and N.F. acknowledge Research Grants funded jointly by Ministero
dell'Istruzione, dell'Universit\`a e della Ricerca (MIUR), by
Universit\`a di Torino and by Istituto Nazionale di Fisica Nucleare
within the {\sl Astroparticle Physics Project} (MIUR contract number:
PRIN 2008NR3EBK; INFN grant code: FA51). S.S. acknowledges support by
the National Research Foundation of Korea (NRF) with a grant funded by
the Korea government (MEST) no. 2011-0024836. N.F. acknowledges
support of the spanish MICINN Consolider Ingenio 2010 Programme under
grant MULTIDARK CSD2009- 00064.

\end{document}